\documentclass[aps,prd,
onecolumn,groupedaddress,showpacs,nofootinbib,amssymb]
{revtex4-2}

\usepackage[dvips]{graphicx}
\usepackage{amssymb}
\usepackage{amsmath}
\usepackage{graphicx,color,comment,physics}
\usepackage{amsfonts}
\usepackage{bm}

\allowdisplaybreaks[4]

\begin{document}

\tolerance=5000

\def\be{\begin{align}}
\def\ee{\end{align}}
\newcommand\nn{\nonumber \\}
\newcommand\e{\mathrm{e}}
\newcommand{\correct}[1]{\color{red}#1\color{black} }

\title{Compact Star in General $F(R)$ Gravity: \\ 
Inevitable Degeneracy Problem and Non-Integer Power Correction
} 

\author{Kota Numajiri$^1$, Taishi Katsuragawa$^2$, and Shin'ichi~Nojiri$^{1,3}$}
\affiliation{
$^1$Department of Physics, Nagoya University, Nagoya 464-8602, Japan\\
$^2$ Institute of Astrophysics, Central China Normal University, Wuhan 430079, China\\
$^3$Kobayashi-Maskawa Institute for the Origin of Particles and the
Universe, Nagoya University, Nagoya 464-8602, Japan
}

\begin{abstract}
We investigate a compact star in the general $F(R)$ gravity.
Developing a novel formulation in the spherically symmetric and static space-time with the matter, we confirm that an arbitrary relation between the mass $M$ and the radius $R_s$ of the compact star can be realized by adjusting the functional form of $F(R)$.
Such a degeneracy with a choice of the equation of state (EOS) suggests that only mass-radius relation is insufficient to constrain the $F(R)$ gravity.
Furthermore, by solving the differential equation for $\left. \frac{dF(R)}{dR}\right|_{R=R(r)}$ near and inside the surface of the compact star with the polytropic EOS,
the boundary condition demands a weak curvature correction to the Einstein gravity could be non-integer power of the scalar curvature,
which gives a stringent constraint on the functional form of $F\left(R\right)$. 
This consequence follows that the equation of motion in $F(R)$ gravity includes the second-order derivative of Ricci scalar $R$, and thus, it is applicable to general $F(R)$ models.
\end{abstract}

\maketitle

\section{Introduction \label{SecI}}

Astrophysical observations confirmed neutron stars whose masses are about twice the Solar mass in 2010 and 2013~\cite{Demorest:2010bx, Antoniadis:2013pzd}.
Such masses were beyond the upper limit and out of the mass-radius relation of neutron stars implied from the hadron physics with the Einstein gravity at that time. 
After that, it has been shown that the quark-matter cores could modify the mass-radius relation and solve this inconsistency~\cite{Annala:2019puf}. 
On the other hand, the modified gravity theories (see the reviews, 
Refs.~\cite{Nojiri:2006ri, Faraoni:2010pgm, Nojiri:2010wj, Capozziello:2011et, Nojiri:2017ncd}, for example) could also refine the relation and solve this problem (see, Refs.~\cite{Arapoglu:2010rz, Astashenok:2013vza, Capozziello:2015yza, Katsuragawa:2015lbl, Feng:2017hje, Yamazaki:2018szv, AparicioResco:2016xcm, Yazadjiev:2018xxk, Feola:2019zqg, Nobleson:2021itp}, for example).
That is, the model-dependence of the equation of state (EOS) and modification of gravity degenerate to each other, and thus, the mass-radius relation is not conclusive enough to separately constrain hadron physics and modified gravity. 
This fact has suggested that there are some possibilities to verify or constrain the modified gravity theories
by the observation in LIGO, Virgo, KAGRA, and DECIGO of the gravitational waves generated 
in the mergers of neutron star binaries or a neutron star and black hole binaries~\cite{LIGOScientific:2017vwq, LIGOScientific:2021qlt}. 

In the study of compact stars in modified gravity, $F(R)$ gravity has been intensively investigated so far, where the action is written by an appropriate function of the scalar curvature $R$.
The existing studies formulated the modified Tolman-Oppenheimer-Volkov (TOV) equations for the general $F(R)$ function, and they have revealed the mass-radius relation which can explain the massive neutron stars with concrete $F(R)$ models and EOS.
Those studies indeed show synergistic constraints on the $F(R)$ gravity models from both cosmology and astrophysics.~\cite{Oikonomou:2021iid, Odintsov:2021nqa,Odintsov:2021qbq}. 
On the other hand, the degeneracy problem still exists, and one naively expects that the functional degree of freedom for the $F(R)$ function degenerates to that of the EOS.
It is analogous to the well-known reconstruction technique for cosmology~\cite{Nojiri:2005sr, Capozziello:2006dj, Cognola:2006sp, Nojiri:2009kx},
where an arbitrary time-evolution of scale factor can be reproduced by choosing $F(R)$ function.
Furthermore, the field equation in $F(R)$ gravity includes the fourth-order of derivative,
although the Einstein equation does the second order.
Thus, one needs to impose additional boundary constraints to solve the modified TOV equations numerically.
There are lots of arguments about the boundary condition;
for example, 
one can solve ordinary junction conditions for the Ricci scalar~\cite{Feng:2017hje} or impose a condition that the Ricci scalar takes the same value as that in general relativity~\cite{Arapoglu:2010rz}, and the scalar degree of freedom in $F(R)$ gravity may allow the nontrivial exterior solution~\cite{Nashed:2020mnp}.
However, one also expects that 
the boundary condition at the surface of compact star potentially constrain the $F(R)$ function,
assuming the Schwarzschild space-time as the exterior geometry.

In this paper, we investigate a spherically symmetric and static space-time in the general $F(R)$ gravity~\cite{Capozziello:2002rd, Nojiri:2003ft} with matters, which describes the compact stars such as the neutron stars. 
Adopting a new expression of modified TOV equations,
we demonstrate that the arbitrary mass-radius relation of the compact star can be realized by adjusting the functional form of $F(R)$ in an analytical way. 
Moreover, we investigate the modified TOV equations around the surface of the compact star perturbatively, which corresponds to the weak curvature expansion, and analyze the additional boundary condition from the higher-order derivatives with respect to the radial coordinate $r$.
The analyses tell us that the correction to the Einstein gravity could be non-integer power of the scalar curvature if we assume the polytrope-type EOS.  

In the next section, we derive the field equations for the spherically symmetric and static space-time in the general $F(R)$ gravity with the matter. 
In Section \ref{SecIII}, we formulate the differential equation for $\frac{dF\left(R\right)}{dR}$ as a function of the radial coordinate $r$, 
which corresponds to one of the modified TOV equations,
and discuss how we determine $F(R)$ function for a given density profile $\rho=\rho(r)$ of the compact star. 
In Section \ref{SecIV}, we explicitly solve the differential equation near the surface of the compact star and find the non-integer power of the curvature as a correction to the Einstein gravity. 
The last section is devoted to summary and discussions.  

\section{$F(R)$ Gravity in Spherically Symmetric Space-time \label{SecII}}

In this section, we briefly review the $F(R)$ gravity and give the modified TOV equations in the spherically symmetric and static space-time.
The action of the $F(R)$ gravity is given by replacing the scalar curvature $R$ in the Einstein-Hilbert action which is,
\begin{align}
\label{JGRG6}
S_\mathrm{EH}=\int d^4 x \sqrt{-g} \left(
\frac{R}{2\kappa^2} + \mathcal{L}_\mathrm{matter} \right)\, ,
\end{align}
by using some appropriate function of the scalar curvature, as follows,
\begin{align}
\label{JGRG7}
S_{F(R)}= \int d^4 x \sqrt{-g} \left(
\frac{F(R)}{2\kappa^2} + \mathcal{L}_\mathrm{matter} \right)\, .
\end{align}
In Eqs.~\eqref{JGRG6} and \eqref{JGRG7},
$\mathcal{L}_\mathrm{matter}$ is the matter Lagrangian density.
Variation of Eq.~\eqref{JGRG7} with respect to the metric leads to the equation of motion for the $F(R)$ gravity theory as follows,
\begin{align}
\label{JGRG13}
G^F_{\mu\nu} \equiv \frac{1}{2}g_{\mu\nu} F - R_{\mu\nu} F_R - g_{\mu\nu} \Box F_R
+ \nabla_\mu \nabla_\nu F_R
= - \kappa^2 T_{\mu\nu}\, .
\end{align}
Here $F_R \equiv \frac{dF(R)}{dR}$ and $T_{\mu\nu}$ is the energy momentum tensor of the matters. 
By using the identity for the scalar function $f$, 
$\nabla_\mu \Box f = - R_{\mu\nu} \nabla^\nu f + \Box \nabla_\mu f$, 
and the Bianchi identity $\nabla^\mu R_{\mu\nu} - \frac{1}{2} \nabla_\nu R =0$, 
we can show that the r.h.s. is consistent with the conservation law 
$\nabla^\mu T_{\mu\nu}=0$; that is,
\begin{align}
\label{JGRG13B}
\nabla^\mu \left( \frac{1}{2}g_{\mu\nu} F - R_{\mu\nu} F_R - g_{\mu\nu} \Box F_R
+ \nabla_\mu \nabla_\nu F_R \right)=0 \, .
\end{align}

For the spherically symmetric and static space-time, 
the metric ansatz is defined as 
\begin{align}
\label{GBiv}
ds^2 = \sum_{\mu,\nu=t,r,\theta,\phi} g_{\mu\nu} dx^\mu dx^\nu =& - \e^{2\nu (r)} dt^2 + \e^{2\lambda (r)} dr^2 
+ r^2 \sum_{i,j=\theta,\phi} {\bar g}_{ij} dx^i dx^j \, , \\
\sum_{i,j=\theta,\phi} {\bar g}_{ij} dx^i dx^j =&\, d\theta^2 + \sin^2\theta \, d\phi^2 \, .
\end{align}
We introduce the perfect fluid as the matter,
where the non-vanishing components of the energy-momentum tensor $T_{\mu\nu}$ are given by 
\begin{align}
\label{FRN1}
T_{tt} = - g_{tt} \rho = \e^{2\nu} \rho 
\, , \quad 
T_{rr} = g_{rr} p = \e^{2\lambda} p 
\, , \quad 
T_{ij} = g_{ij} p = r^2 \bar{g}_{ij} p 
\, .
\end{align}
Here $\rho$ and $p$ are the total energy density and the pressure of the matters, respectively, and satisfy the following conservation law,  
\begin{align}
\label{FRN2}
0 = \nabla^\mu T_{\mu r} = \nu' \left( \rho + p \right) + \frac{dp}{dr} \, .
\end{align}
Here $\rho$ and $p$ only depend on the radial coordinate $r$. 
Other components of the conservation law are trivially satisfied. 
If the EOS $\rho=\rho(p)$ is given, 
Eq.~\eqref{FRN2} can be integrated as 
\begin{align}
\label{FRN3}
\nu (r) = - \int^r dr \frac{\frac{dp}{dr}}{\rho + p} 
= - \int^{p(r)}\frac{dp}{\rho(p) + p} \, .
\end{align}
Then, the non-vanishing components of Eq.~\eqref{JGRG13} are as follows:
\begin{align}
\label{FRN4}
 - \kappa^2 \rho =&  
 - \frac{1}{2} F - \e^{- 2 \lambda} \left[
\nu'' + \left(\nu' - \lambda'\right)\nu' + \frac{2\nu'}{r}\right] F_R 
+ \e^{ -2\lambda} \left[ F_R'' + \left( - \lambda' + \frac{2}{r} \right) F_R' \right]  \, ,\\
\label{FRN5}
 - \kappa^2 p =&\, 
\frac{1}{2} F + \e^{ -2\lambda} \left[ \nu'' + \left(\nu' - \lambda'\right)\nu' 
 - \frac{2 \lambda'}{r} \right] F_R - \e^{ -2\lambda} 
\left( \nu' + \frac{2}{r} \right) F_R' \, ,\\
\label{FRN6}
 - \kappa^2 p =&\, 
\frac{1}{2} F - \frac{1}{r^2} \left\{ 1 + \left[ - 1 - r \left(\nu' 
 - \lambda' \right)\right] \e^{-2\lambda}\right\} F_R 
 - \e^{-2\lambda} \left[ F_R'' + \left( \nu' - \lambda' + \frac{1}{r} \right) F_R' \right] \, .
\end{align}
Note that using Eq.~\eqref{JGRG13B}, 
one finds Eqs.~\eqref{FRN4}, \eqref{FRN5}, and \eqref{FRN6} reproduces the conservation law in Eq.~\eqref{FRN2}.

\section{Reconstruction of $F(R)$ via the mass-radius relation \label{SecIII}}

In the Einstein gravity, the TOV equations determine the density profile $\rho(r)$ of compact stars: 
\begin{align}
    m'(r)=&\, 4 \pi r^{2} \rho(r) 
    \, , \\
    p'(r)=& -\left[ \rho(r)+p(r) \right] \frac{m(r) + 4 \pi r^{3} p(r)}{r\left[r-2 m(r)\right]}
    \, ,
\end{align}
where $m(r) = \frac{r}{2} \left( 1-\e^{-2\lambda(r)} \right)$ is the Schwarzschild mass. 
For a given EOS $p=p(\rho)$ inside the star, one can solve the above differential equations for $\rho(r)$ and $m(r)$,
and computing the Schwarzschild mass $m(r)$ for a number of initial conditions (central density) of density profile $\rho(r)$ gives us the mass-radius relation.
It is known that in the Einstein gravity, 
the EOS and the mass-radius relation (or the density profile) have one-to-one correspondence via inverse-TOV method ~\cite{Lindblom1992}.
On the other hand, in the $F(R)$ gravity, 
the two degrees of freedom from the EOS and the functional form of $F(R)$ degenerate to each other.
Hence we can determine one of these three (the EOS, the density profile, and $F(R)$ function) by using the other two.
For instance, when we calculate the mass-radius relation (thus, the density profile), 
we have to specify the EOS and the function $F(R)$. 
Solving two of Eqs.~\eqref{FRN4}-\eqref{FRN6} with Eq.~\eqref{FRN2}, the existing works demonstrated the mass-radius relation in $F(R)$ gravity~\cite{Astashenok:2013vza, Capozziello:2015yza, AparicioResco:2016xcm, Arapoglu:2010rz, Feola:2019zqg}.

In this section, we consider the inverse problem; we try to find the functional form of $F(R)$ with the given EOS and density profile. 
Main goal is deriving the differential equation of $F_R$ with respect to the radial coordinate $r$ for a given EOS $p=p(\rho)$ and density profile of $\rho=\rho(r)$.
Integrating the density profile gives us the relation between the mass $M$ and the radius $R_s$ of the compact star in a way similar to the ordinary TOV equations in the Einstein gravity. 
However, as we will see later, for the arbitrary relation between the mass $M$ and the radius $R_s$ of the compact star, the $F(R)$ model can be reconstructed to reproduce a desired $M$-$R_s$ relation for the arbitrary EOS. 

Combining Eqs.~\eqref{FRN4} and \eqref{FRN5}, Eqs.~\eqref{FRN4} and \eqref{FRN6}, and Eqs.~\eqref{FRN5} and \eqref{FRN6}, 
we obtain the following three equations, respectively:
\begin{align}
\label{FRN8}
- \kappa^2 \left(\rho + p \right) 
= &
- \frac{2\left(\nu' + \lambda'\right)}{r} \e^{- 2 \lambda} F_R 
+ \e^{ -2\lambda} \left[ F_R'' - \left( \nu'  + \lambda' \right) F_R' \right] 
\, , \\
\label{FRN9}
- \kappa^2 \left( \rho + p \right)
= &
- \left\{ \frac{1}{r^2} + \e^{- 2 \lambda} \left[ 
\nu'' + \left(\nu' - \lambda'\right)\nu' + \frac{\nu'+\lambda'}{r}
 - \frac{1}{r^2} \right] \right\} F_R 
 - \e^{-2\lambda} \left( \nu' - \frac{1}{r} \right) F_R' 
\, , \\
\label{FRN10}
0 =& \left\{ \frac{1}{r^2} + \e^{ -2\lambda} \left[ \nu'' + \left(\nu' - \lambda'\right)\nu' 
 - \frac{1}{r^2} - \frac{\nu' + \lambda'}{r} \right] \right \} F_R 
+ \e^{-2\lambda} \left[ F_R'' + \left( - \lambda' - \frac{1}{r} \right) F_R' \right]
\, . 
\end{align}
Since one of the above three equations is redundant by the conservation law in Eq.~\eqref{FRN2},
the analysis below focuses on Eqs.~\eqref{FRN8} and \eqref{FRN9}.
In usual TOV-like problems, these are the dynamical equations to determine the metric components $\lambda$ and $\nu$, and by combining them with the EOS and Eq.~\eqref{FRN2}, the density profile $\rho(r)$ is obtained. 
However, since the $F(R)$ function is to be solved here, we should construct the differential equation for $F(R)$ deleting unknown functions such as $\lambda$. 
For this purpose, we now define a new variable $N(r)\equiv \e^{-2\nu - 2\lambda}$ and delete $\lambda$ in Eq.~\eqref{FRN8}, 
to find the differential equation for $N(r)$, 
\begin{align}
\label{FRN12}
N' \left( \frac{F_R}{r} + \frac{1}{2} F_R' \right) + N F_R'' 
= - \e^{-2\nu} \kappa^2 \left(\rho + p \right)\, , 
\end{align}
which can be integrated with respect to $N$, as follows, 
\begin{align}
\label{FRN13}
N=& -\kappa^2 \exp \left( - \int_{R_s}^r dr_1 \frac
{\frac{d^2 F_R \left(R\left(r_1\right)\right)}{d {r_1}^2}}
{\frac{F_R\left(R\left(r_1\right)\right)}{r_1} 
 + \frac{1}{2} \frac{d F_R \left(R\left(r_1\right)\right)}{d {r_1}}} \right) \nn
& \times \left( 
C + 
\int_{R_s}^r dr_2 \exp \left( \int_{R_s}^{r_2} dr_3 \frac
{\frac{d^2 F_R\left(R\left(r_3\right)\right)}{dr_3^2}}
{\frac{F_R\left(R\left(r_3\right)\right)}{r_3} 
 + \frac{1}{2} \frac{d F_R\left(R\left(r_3\right)\right)}{dr_3}} \right)
\frac{ \e^{-2\nu\left( r_2 \right)} \left(\rho\left( r_2 \right) 
+ p\left( r_2 \right) \right)}
{\frac{F_R\left(R\left(r_2\right)\right)}{r_2} 
 + \frac{1}{2} \frac{d F_R\left(R\left(r_2\right)\right)}{dr_2}} \right) \, ,
\end{align}
Here $C$ is a constant of integration. 
Rewriting Eq.~\eqref{FRN9} in terms of $N$, 
\begin{align}
\label{FRN14}
 - \left[ N \left( \nu'' + 2 {\nu'}^2 - \frac{1}{r^2} \right) 
+ N'\left( \frac{\nu'}{2} -  \frac{1}{2r} \right) \right] F_R 
 - N \left( \nu' - \frac{1}{r} \right) F_R' 
= \e^{-2\nu} \left[ - \kappa^2 \left( \rho + p \right) 
+ \frac{1}{r^2} F_R\right] \, ,
\end{align}
and substituting Eq.~\eqref{FRN13} into the above, we obtain 
\begin{align}
\label{FRN16}
& \int_{R_s}^r dr_2 \exp \left( C + \int_{R_s}^{r_2} dr_3 
\frac{\frac{d^2 F_R \left(R\left(r_3\right)\right)}{d r_3^2}} 
{\frac{F_R \left(R\left(r_3\right)\right)}{r_3} 
+ \frac{1}{2} \frac{d F_R \left(R\left(r_3\right)\right)}{d r_3}}
\right)
\frac{ \e^{-2\nu\left( r_2 \right)} \kappa^2 \left(\rho\left( r_2 \right) 
+ p\left( r_2 \right) \right)}
{\frac{F_R \left(R\left(r_2\right)\right)}{r_2} 
 + \frac{1}{2} \frac{d F_R \left(R\left(r_2\right)\right)}{d r_2}} \nn
=& \left[ \left( \nu'' + 2 {\nu'}^2 - \frac{1}{r^2}\right) F_R
 - \left( \frac{\nu'}{2} -  \frac{1}{2r} \right) 
 \frac{F_RF_R''}{\frac{F_R}{r} 
+ \frac{1}{2} F_R' } + \left( \nu' - \frac{1}{r} \right)
F_R' \right]^{-1} \nn
& \times \e^{-2\nu} \left\{ - \kappa^2 \left( \rho + p \right) \left[ 1 
+ \frac{ \left(\frac{\nu'}{2} -  \frac{1}{2r}\right)F_R}{\frac{F_R}{r} 
+ \frac{1}{2} F_R' } \right]
+ \frac{1}{r^2} F_R\right\} 
\exp \left( \int_{R_s}^r dr_1 
\frac{\frac{d^2 F_R \left(R\left(r_1\right)\right)}{d r_1^2}} 
{\frac{F_R \left(R\left(r_1\right)\right)}{r_1} 
 + \frac{1}{2} \frac{d F_R \left(R\left(r_1\right)\right)}{d r_1}}
\right) \, .
\end{align}
By differentiating the above expression with respect $r$, we obtain, 
\begin{align}
\label{FFF1}
& \kappa^2 \left(\rho + p \right)\left\{ \frac{1}{\frac{F_R}{r} + \frac{1}{2} F_R'} 
\left[ \left( \nu'' + 2 {\nu'}^2 - \frac{1}{r^2}\right) F_R
 - \left( \nu' -  \frac{1}{r} \right) 
\frac{F_RF_R''} {2 \left( \frac{F_R}{r} 
+ \frac{1}{2} F_R' \right) } 
+ \left( \nu' - \frac{1}{r} \right)F_R' \right]^2 \right. \nn 
& - \left[ \left( \nu''' + 4 \nu' \nu'' + \frac{2}{r^3}\right) F_R
+ 2 \left( \nu'' + {\nu'}^2 \right) F_R'
+ \left( \nu' - \frac{1}{r} \right)F_R'' 
 - \left( \nu'' +  \frac{1}{r^2} \right) \frac{F_RF_R''} {2 \left( \frac{F_R}{r} 
+ \frac{1}{2} F_R' \right) } \right. \nn
& \left. - \left( \nu' -  \frac{1}{r} \right) \frac{1} {2 \left( \frac{F_R}{r} 
+ \frac{1}{2} F_R' \right)^2 } \left[ - \left(
 - \frac{F_R}{r^2}+ \frac{1}{2} F_R''\right) F_R F_R'' 
+ \frac{1}{2} {F_R'}^2 F_R'' + \left(\frac{F_R}{r} 
+ \frac{1}{2} F_R'\right) F_R F_R''' \right] \right] \nn
& \times \left[ 1 + \frac{ \left(\nu' -  \frac{1}{r}\right)F_R}{2 \left( \frac{F_R}{r} 
+ \frac{1}{2} F_R' \right)} \right] \nn
& + \left[ - 2 \nu'  
+ \frac{ \left(\nu'' - 2 {\nu'}^2 + \frac{2\nu'}{r} + \frac{1}{r^2}\right)F_R + \left(\nu'-\frac{1}{r}\right)F_R' + 2 F_R''}
{2\left( \frac{F_R}{r} + \frac{1}{2} F_R' \right) }
 - \frac{F_R \left(\nu' -  \frac{1}{r}\right) \left(\frac{F_R'}{r} 
 - \frac{F_R}{r^2} - \frac{1}{2} F_R''\right)}{2 \left( \frac{F_R}{r} 
+ \frac{1}{2} F_R' \right)^2} \right] \nn
& \left. \times \left[ \left( \nu'' + 2 {\nu'}^2 - \frac{1}{r^2}\right) F_R
 - \left(\nu' -  \frac{1}{r} \right) 
\frac{F_RF_R''} {2 \left( \frac{F_R}{r} 
+ \frac{1}{2} F_R' \right) } 
+ \left( \nu' - \frac{1}{r} \right)F_R' \right] \right\} \nn
=& - \left\{ \left( \nu''' + 4 \nu' \nu'' + \frac{2}{r^3}\right) F_R
+ 2 \left( \nu'' + {\nu'}^2 \right) F_R'
+ \left( \nu' - \frac{1}{r} \right)F_R'' 
 - \left( \nu'' +  \frac{1}{r^2} \right) \frac{F_RF_R''} {2 \left( \frac{F_R}{r} 
+ \frac{1}{2} F_R' \right) } \right. \nn
& \left. - \left( \nu' -  \frac{1}{r} \right) \frac{1} {2 \left( \frac{F_R}{r} 
+ \frac{1}{2} F_R' \right)^2 } \left[ - \left(
 - \frac{F_R}{r^2}+ \frac{1}{2} F_R''\right) F_R F_R'' 
+ \frac{1}{2} {F_R'}^2 F_R'' + \left(\frac{F_R}{r} 
+ \frac{1}{2} F_R'\right) F_R F_R''' \right] \right\} \frac{1}{r^2} F_R \nn
& + \left[ \left( \nu'' + 2 {\nu'}^2 - \frac{1}{r^2}\right) F_R
 - \left(\nu' -  \frac{1}{r} \right) 
\frac{F_RF_R''} {2 \left( \frac{F_R}{r} 
+ \frac{1}{2} F_R' \right) } 
+ \left( \nu' - \frac{1}{r} \right)F_R' \right] \nn
& \times  \left\{ - \frac{2\nu' }{r^2} F_R - \frac{2}{r^3} F_R + \frac{1}{r^2} F_R' 
+ \frac{F_R F_R''}{r^2 \left( \frac{F_R}{r} + \frac{1}{2} F_R'\right)}
 - \kappa^2 \frac{d \left( \rho + p \right)}{dr} \left[ 1 
+ \frac{ \left(\nu' -  \frac{1}{r}\right)F_R}{2 \left( \frac{F_R}{r} 
+ \frac{1}{2} F_R' \right) } \right] \right\} 
\, .
\end{align}
Although the above expression seems complicated, Eq.~\eqref{FFF1} represents a differential equation of $F_R$ 
with respect to the radial coordinate $r$ if $\nu$, $\rho$, and $p$ are given by the functions of $r$. 

Once the equation of the state $p=p(\rho)$ and the density profile $\rho (r)$ are given, 
$p$ becomes a function of the radial coordinate, $p = p (\rho(r))$,
and Eq.~\eqref{FRN3} determines the $r$ dependence of $\nu = \nu(r)$.
Using two profiles $p=p(r)$ and $\nu=\nu(r)$, 
one can obtain the mass-radius relation of the compact star
by demanding that $\nu$ is smoothly connected to the Schwarzschild one $\e^{2\nu}=1-2M/r$ (or by spherically integrating $\rho(r)$) as the boundary condition. 
Note that we have not used the details of the gravitational theory, which is to be derived as a solution of Eq.~\eqref{FFF1}, and it shows the difference from the usual TOV configuration. 

Eq.~\eqref{FFF1} provides us with $F_R$ as a function of the radial coordinate $r$, $F_{R} = F_{R}(r)$. 
By substituting the obtained expression of $F_R (r)$ into Eq.~\eqref{FRN13},
we find $\lambda(r)$ with $N= \e^{-2\nu - 2\lambda}$ and known $\nu(r)$.
Furthermore, substituting $\nu=\nu(r)$, $\lambda=\lambda(r)$ into the scalar curvature $R$ in Eq.~\eqref{curvaturesS}, 
we find the scalar curvature $R$ as a function of $r$.
Deleting $r$ by the inverse function $r=r(R)$, we reach the explicit form of $F_R(R)$ and therefore $F(R)$. 
The above procedure implies that the functional form of $F(R)$ can be reconstructed from the EOS and the density profile of a compact star.

As was demonstrated, if the EOS and the density profile are given, the mass-radius relation and the functional form of $F(R)$ can be computed simultaneously. 
In other words, this result concludes that for the arbitrary mass-radius relation (or the density profile) of the compact star for the arbitrary EOS, therefore, we can find the model of gravity theory reproducing the $M$-$R_s$ relation. 

\section{Constraints from Boundary Condition at Surface of Compact Star \label{SecIV}}

\subsection{Energy-polytrope case}
We have confirmed that the degeneracy issue is inevitable due to the model-dependence of the EOS and $F(R)$ function in the previous section, and thus, one can reconstruct the $F(R)$ function to lead to a given $M$-$R_s$ relation.
However, it does not mean that any functional form of $F(R)$ is allowed.
Because field equations include the second-order derivative of the scalar curvature $R$, the first derivative of the scalar curvature $R'$ must be continuous at the surface $r=R_s$ even if the values of the matter-energy density $\rho$ and pressure $p$ jump at the surface. 
Therefore near the surface, $R$ behaves as 
\begin{align}
\label{sc1}
R \sim R_0 \left( 1 - \frac{r}{R_s} \right)^\alpha 
\, ,
\end{align}
where $\alpha \geq 2$ and $R_0$ is a dimensional constant. 
Because $R$ includes $\lambda$, $\lambda'$, $\nu'$, and $\nu''$ as given in \eqref{curvaturesS}, the continuity of $R'$ suggests the continuities of $\lambda$, $\lambda'$, $\lambda''$, $\nu'$, $\nu''$, and $\nu'''$. 
Note that in the Einstein gravity, we need to require $\nu$ and $\nu'$ to be continuous at the surface to avoid the singularity in the equations; however, one can admit a jump in $\nu''$ and the curvatures at the surface. 
In this section, we investigate a possible form of $F(R)$ near and inside the surface of the compact star constrained by the exterior Schwarzschild geometry.

First of all, it is necessary that the Schwarzschild space-time should be a vacuum solution in $F(R)$ gravity.
Assuming that the Ricci tensor is covariantly constant $R_{\mu \nu} \propto g_{\mu \nu}$ in vacuum, 
one finds the equation of motion Eq.~\eqref{JGRG13} leads to
\begin{align}
0=2F(R) - R F_{R} (R)
\, .
\end{align}
The Schwarzschild-(anti-)de Sitter space-time is an exact solution if the above has a solution.
Note that the effect of cosmological constant, which is responsible for the asymptotic (anti-)de Sitter geometry, 
is negligible at the local scale, and one can use the Schwarzschild space-time as the exterior one of the compact star.
Moreover, to keep as much generality as possible and to perform the calculations in an analytic manner, 
we consider the energy-polytrope, whose EOS is given by 
\begin{align}
\label{polytrope}
p = K \rho^{1 + \frac{1}{n}}
\, ,
\end{align}
with constants $K$ and $n$.
It is known that for the neutron stars, $n$ could take the value $0.5\leq n \leq 1$~\cite{1939isss.book.....C}.
Eq.~\eqref{polytrope} reads with respect to $\rho$,
\begin{align}
\label{polytrope2}
\rho = \tilde K p^{1 + \frac{1}{\tilde n}}\, , \quad 
\tilde K \equiv K^{-\frac{1}{1+\frac{1}{n}}} \, , \quad
\tilde n \equiv \frac{1}{\frac{1}{1+\frac{1}{n}} - 1} 
= - 1 - n \, .
\end{align}
For the energy-polytrope, Eq.~\eqref{FRN3} takes the following form~\footnote{
By changing the variable as $p=C\e^s$ with a constant $C$, 
we can rewrite \eqref{FRN3} as 
\[
\nu(r) =
 - \int^{\ln \frac{p}{C}} \frac{ds}{\tilde K C^{\frac{1}{\tilde n}} \e^{\frac{s}{n}} + 1}
= - \int^{\ln \frac{p}{C}} \frac{\e^{-\frac{s}{\tilde n}}ds}{\tilde K C^{\frac{1}{\tilde n}} 
+ \e^{-\frac{s}{\tilde n}}}
= \nu_0 + \left. \tilde n \ln \left( 1 + {\tilde K}^{-1} \left(C\e^s\right)^{-\frac{1}{\tilde n}} \right) 
\right|_{s=\ln \frac{p}{C}} 
=
\nu_0 - \left(n+1\right) \ln \left( 1 + K^{\frac{1}{1+\frac{1}{n}}} p^{\frac{1}{n+1}} \right) \, . 
\]
},
\begin{align}
\label{FRN3p1B}
\nu (r) = - \int^{p(r)}\frac{dp}{\tilde K p^{1 + \frac{1}{\tilde n}} + p} 
= \nu_0 + \tilde n \ln \left( 1 + {\tilde K}^{-1} p^{-\frac{1}{\tilde n}} \right) \, .
\end{align}
Here $\nu_0$ is a constant of the integration. 

To analyze the boundary condition, the $r$ dependence of density profile near the surface of the star is essential. 
Although the pressure should vanish at the surface $r=R_s$, the convergent behavior is nontrivial.
Using the Taylor expansion around the surface, we assume the following profile for $p=p(r)$ at $r\lesssim R_s$:
\begin{align}
\label{ProG1B}
p(r)\sim p_0 \left( 1 - \frac{r}{R_s} \right)^m \left[ 1 + p_1 \left( 1 - \frac{r}{R_s} \right) + p_2 \left( 1 - \frac{r}{R_s} \right)^2 + \cdots \right] \, ,
\end{align}
where $m$ is a constant determined by the consistency later.
If we assume the energy-polytrope in Eq.~\eqref{polytrope}, 
Eq.~\eqref{FRN3p1B} gives 
\begin{align}
\label{ProG2B}
\nu (r) \sim 
\nu_0 - \left( 1 + n \right) \ln \left\{ 1 + \nu_1 \left( 1 - \frac{r}{R_s} \right)^{\frac{m}{1+ n}} 
\left[ 1 + p_1 \left( 1 - \frac{r}{R_s} \right) + p_2 \left( 1 - \frac{r}{R_s} \right)^2 + \cdots \right]^{\frac{1}{1+ n}}
\right\} \, .
\end{align}
Here, $\nu_{1}$ is defined as 
\begin{align}
\label{PP1}
\nu_1 \equiv K^{\frac{1}{1+\frac{1}{n}}} p_0^{\frac{1}{1+ n}} \, .
\end{align}

Outside the star, following the conventional manner in the Einstein gravity,
we assume that the metric should be given by the Schwarzschild space-time, 
\begin{align}
\label{Sch1}
\nu(r) = \frac{1}{2} \ln \left( 1 - \frac{2M}{r} \right) \, .
\end{align}
The conditions for the continuities of $\nu(r)$ and $\nu'(r)$ at the surface of the star $r=R_s$ indicate $\frac{1}{2} \ln \left( 1 - \frac{2M}{R_s} \right) = \nu_0$ and 
\begin{align}
\label{ProG4}
\frac{\frac{M}{R_s^2}}{1 - \frac{2M}{R_s}}
=& \left\{ \frac{m\nu_1}{R_s}\left( 1 - \frac{r}{R_s} \right)^{\frac{m}{1+ n}-1} 
\left[ 1 + p_1 \left( 1 - \frac{r}{R_s} \right) + p_2 \left( 1 - \frac{r}{R_s} \right)^2 + \cdots \right]^{\frac{1}{1+ n}} \right. \nn
& \left. + \frac{\nu_1}{R_s} \left( 1 - \frac{r}{R_s} \right)^{\frac{m}{1+ n}} 
\left[ 1 + p_1 \left( 1 - \frac{r}{R_s} \right) + p_2 \left( 1 - \frac{r}{R_s} \right)^2 + \cdots \right]^{\frac{1}{1+ n} - 1}
\left[ p_1 + 2 p_2 \left( 1 - \frac{r}{R_s} \right) + \cdots \right] \right\} \nn
& \left. \times \left\{ 1 + \nu_1 \left( 1 - \frac{r}{R_s} \right)^{\frac{m}{1+ n}} 
\left[ 1 + p_1 \left( 1 - \frac{r}{R_s} \right) + p_2 \left( 1 - \frac{r}{R_s} \right)^2 + \cdots \right]^{\frac{1}{1+ n}}\right\}^{-1} \right|_{r=R_s}\, .
\end{align}
In order for the r.h.s. of Eq.~\eqref{ProG4} to be finite and non-vanishing, 
two parameters $n$ and $m$ should satisfy the following relation:
\begin{align}
\label{ProG5}
m=n+1 \, .
\end{align}
Here, the above relation is irrespective of the model of gravity because the EOS~\eqref{polytrope} determines the behavior of the pressure $p$ around the surface.

In case of Eq.~\eqref{ProG5}, Eq.~\eqref{ProG4} is reduced to a simple form:
\begin{align}
\label{ProG6B}
\frac{\frac{M}{R_s^2}}{1 - \frac{2M}{R_s}}
= \frac{\left(n+1 \right) \nu_1}{R_s}
= \frac{n+1}{R_s} K^{\frac{1}{1+\frac{1}{n}}} p_0^{ \frac{1}{1+ n}} \, .
\end{align}
That is, 
\begin{align}
\label{ProG7B}
\frac{M}{R_s}=\frac{\left(n+1 \right) K^{\frac{1}{1+\frac{1}{n}}} p_0^{\frac{1}{1+ n}}}
{1+2\left(n+1 \right) K^{\frac{1}{1+\frac{1}{n}}} p_0^{\frac{1}{1+ n}}} 
= \frac{1}{2} \left[ 1 - \frac{1}{1+2\left(n+1 \right) K^{\frac{1}{1+\frac{1}{n}}} p_0^{\frac{1}{1+ n}}} \right] \, ,
\end{align}
which constrains the compactness of the star written in the mass $M$ and the radius $R_s$ as follows,
\begin{align}
\label{FRsrfc0}
0< \frac{M}{R_s} < \frac{1}{2} \, .
\end{align}
We emphasize that the above constraint follows from the behavior near and inside the surface and 
does not depend on the details of the inner structure in the star. 

When Eq.~\eqref{ProG5} is satisfied, Eq.~\eqref{ProG2B} has the following form, 
\begin{align}
\label{ProG8B}
\nu (r) \sim&\,  
\nu_0 - \left( 1 + n \right) \ln \left\{ 1 + \nu_1 \left( 1 - \frac{r}{R_s} \right) 
\left[ 1 + p_1 \left( 1 - \frac{r}{R_s} \right) + p_2 \left( 1 - \frac{r}{R_s} \right)^2 + \cdots \right]^{\frac{1}{1+ n}}
\right\} \, . \nn
\nu' (r) \sim & \left\{ \frac{\left( 1 + n \right) \nu_1}{R_s} \left[ 1 + p_1 \left( 1 - \frac{r}{R_s} \right) 
+ p_2 \left( 1 - \frac{r}{R_s} \right)^2 + \cdots \right]^{\frac{1}{1+ n}} \right. \nn
& \left. + \frac{\nu_1}{R_s} \left( 1 - \frac{r}{R_s} \right) 
\left[ 1 + p_1 \left( 1 - \frac{r}{R_s} \right) + p_2 \left( 1 - \frac{r}{R_s} \right)^2 + \cdots \right]^{\frac{1}{1+ n} - 1}
\left[ p_1 + 2 p_2 \left( 1 - \frac{r}{R_s} \right) + \cdots \right] \right\} \nn
& \times \left\{ 1 + \nu_1 \left( 1 - \frac{r}{R_s} \right) 
\left[ 1 + p_1 \left( 1 - \frac{r}{R_s} \right) + p_2 \left( 1 - \frac{r}{R_s} \right)^2 + \cdots \right]^{\frac{1}{1+ n}}\right\}^{-1} \, , \nn
\nu'' (r) \sim & \left\{ - \frac{2\nu_1}{R_s^2} 
\left[ 1 + p_1 \left( 1 - \frac{r}{R_s} \right) + p_2 \left( 1 - \frac{r}{R_s} \right)^2 + \cdots \right]^{\frac{1}{1+ n} - 1}
\left[ p_1 + 2 p_2 \left( 1 - \frac{r}{R_s} \right) + \cdots \right] \right. \nn
& + \frac{n \nu_1}{\left( 1 + n \right) R_s^2} \left( 1 - \frac{r}{R_s} \right) 
\left[ 1 + p_1 \left( 1 - \frac{r}{R_s} \right) + p_2 \left( 1 - \frac{r}{R_s} \right)^2 + \cdots \right]^{\frac{1}{1+ n} - 2}
\left[ p_1 + 2 p_2 \left( 1 - \frac{r}{R_s} \right) + \cdots \right]^2 \nn 
& \left. - \frac{2\nu_1 p_2}{R_s^2} \left( 1 - \frac{r}{R_s} \right)
\left[ 1 + p_1 \left( 1 - \frac{r}{R_s} \right) + p_2 \left( 1 - \frac{r}{R_s} \right)^2 + \cdots \right]^{\frac{1}{1+ n} - 1}
+ \cdots \right\} \nn
& \times \left\{ 1 + \nu_1 \left( 1 - \frac{r}{R_s} \right) 
\left[ 1 + p_1 \left( 1 - \frac{r}{R_s} \right) + p_2 \left( 1 - \frac{r}{R_s} \right)^2 + \cdots \right]^{\frac{1}{1+ n}}\right\}^{-1} \nn
& + \frac{1}{1 + n} \left\{ \frac{\left( 1 + n \right) \nu_1}{R_s} \left[ 1 + p_1 \left( 1 - \frac{r}{R_s} \right) 
+ p_2 \left( 1 - \frac{r}{R_s} \right)^2 + \cdots \right]^{\frac{1}{1+ n}} \right. \nn
& \left. + \frac{\nu_1}{R_s} \left( 1 - \frac{r}{R_s} \right) 
\left[ 1 + p_1 \left( 1 - \frac{r}{R_s} \right) + p_2 \left( 1 - \frac{r}{R_s} \right)^2 + \cdots \right]^{\frac{1}{1+ n} - 1}
\left[ p_1 + 2 p_2 \left( 1 - \frac{r}{R_s} \right) + \cdots \right] \right\}^2 \nn
& \times \left\{ 1 + \nu_1 \left( 1 - \frac{r}{R_s} \right) 
\left[ 1 + p_1 \left( 1 - \frac{r}{R_s} \right) + p_2 \left( 1 - \frac{r}{R_s} \right)^2 + \cdots \right]^{\frac{1}{1+ n}}\right\}^{-2} \, , \nn
\end{align}
If we further impose the continuity $\nu''(r)$ at the surface of the star $r=R_s>2M$, we find
\begin{align}
\label{ProG9B}
 - \frac{\frac{2M}{R_s^3}}{1 - \frac{2M}{R_s}} 
 - \frac{\frac{2M^2}{R_s^4}}{\left(1 - \frac{2M}{R_s}\right)^2} 
= - \frac{2\nu_1 p_1}{R_s^2} + \frac{\left( 1 + n \right) \nu_1^2}{R_s^2} \, ,
\end{align} 
which can be solved with respect to $p_1$, as follows, 
\begin{align}
\label{FRsrfc1B}
p_1 =  \frac{\left(1+n\right)\nu_1}{2} + \frac{1}{\nu_1} \left[ \frac{\frac{M}{R_s}}{1 - \frac{2M}{R_s}} 
+ \frac{\frac{M^2}{R_s^2}}{\left(1 - \frac{2M}{R_s}\right)^2} \right] \, .
\end{align}
Further by requiring the continuity of $\nu'''$, the parameter $p_2$ in Eq.~\eqref{ProG1B} can be determined as in Eq.~\eqref{FRsrfc1B}. 
The above consequence clarifies the difference between the Einstein gravity and $F(R)$ gravity, originating from the higher derivatives in the field equations.

Finally, we investigate the possible constraint on the $F(R)$ model from the continuity conditions by using Eq.~\eqref{FRN12}.
Because $N\equiv \e^{-2\nu - 2\lambda}=1$ in the Schwarzschild space-time outside the star, 
the continuity of $\lambda$, $\lambda'$, $\lambda''$, $\nu'$, $\nu''$, and $\nu'''$ tell 
$N=1$ and $N'=N''=0$, which tells 
\begin{align}
\label{N}
N = 1 + \mathcal{O}\left( \left( 1 - \frac{r}{R_s} \right)^{N_0} \right) \, , \quad N_0>2 \, .
\end{align}
At the leading order, if $F(R)$ function describes the correction to the Einstein gravity as
\begin{align}
\label{FRp1}
F_R \sim F_0 + F_1 \left(\frac{R}{R_0}\right)^\beta \sim F_0 + F_1 \left( 1 - \frac{r}{R_s} \right)^{\alpha\beta} \, .
\end{align}
For the energy-polytropic EOS \eqref{polytrope}, Eqs.~\eqref{ProG1B} and \eqref{ProG5} conclude 
\begin{align}
\label{FRp4}
p \sim p_0 \left( 1 - \frac{r}{R_s} \right)^{n+1}\, , \quad 
\rho \sim \tilde K p_0^\frac{n}{n+1} \left( 1 - \frac{r}{R_s} \right)^n \, ,
\end{align}
and $\rho + p$ behaves as 
\begin{align}
\label{prho}
\rho + p \sim \rho \sim \left( 1 - \frac{r}{R_s} \right)^n \, .
\end{align}
By using Eqs.~\eqref{N} and \eqref{FRp1}, we find that the first term in the l.h.s. of Eq.~\eqref{FRN12} 
cannot balance with the r.h.s.  of Eq.~\eqref{FRN12} and therefore the second term in the l.h.s. 
must balance with the r.h.s., which tells 
\begin{align}
\label{FRp3}
\alpha \beta - 2 = n \, .
\end{align}
The power $\beta$ of the scalar curvature which characterizes the correction term to the Einstein gravity is thus determined,
\begin{align}
\label{FRp5}
\beta = \frac{n+2}{\alpha}\, ,
\end{align}
and Eq.~\eqref{FRp1} gives us the possible form of $F(R)$ function:
\begin{align}
\label{Nsurface7}
F \left(R\right) \sim \Lambda + F_0 R+ \frac{F_1 R_0}{\beta + 1} \left( \frac{R}{R_0} \right)^{\beta+1} + \cdots\, .
\end{align}
Here $\Lambda$ is a constant of the integration which can be cast as the cosmological constant. 
Eqs.~\eqref{ProG2B} and \eqref{ProG5} show that $\alpha$ is an integer and $\alpha\geq 2$,
moreover $n$ corresponds to the value $0.5\leq n \leq 1$ in the case of neutron stars.
Therefore, the non-integer power of the scalar curvature $R$ can show up in the functional form of $F \left(R\right)$.
We note that the constant $\Lambda$ can spoil our first assumption that the $F(R)$ model possesses the Schwarzschild solution as a vacuum solution, indicating $\Lambda=0$. 
On the other hand, compared with a typical compact star scale, $\Lambda$ is negligible if it takes the value of observed dark energy.

Let us reconsider the other two assumptions.
To show that the weak curvature correction of the $F(R)$ gravity to the Einstein gravity should be non-integer, we have considered the energy-polytrope as in Eq.~\eqref{polytrope} for the equation of state, and we have assumed that the space-time outside of the compact star is the Schwarzschild one. 
The equation of state in the immediate vicinity of the surface can be different from the polytrope one.
Also, due to the scalar mode in the $F(R)$ gravity, the geometry outside the compact star is, in general, different from the Schwarzschild space-time, although the space-time should be asymptotically the Schwarzschild one where $R\to 0$ as in existing works~\cite{Astashenok:2021xpm,Astashenok:2021peo,Astashenok:2020qds,Astashenok:2021btj}.

We should note that if the equation of state near the surface except the very vicinity is given by the energy-polytrope, the geometry inside the star smoothly and monotonically connects to the geometry outside the star. 
When we obtain \eqref{ProG5}, we require $\nu'$ does not diverge nor vanish. 
If $\nu'$ diverges, the inner geometry does not connect with the outer geometry. 
Even if $\nu'$ vanishes, the internal geometry does not connect with the external geometry, either because $\nu'>0$ in the asymptotic Schwarzschild space-time. 
Other possibilities might be that the geometry is oscillating or not monotonically connecting (i.e., there is an inflection point on $R$ inside the surface).
However, the spatial oscillation tells the existence of the tachyon; therefore, the asymptotic Schwarzschild space-time becomes unstable and thus unphysical. 
Also, the existence of an inflection point on the Ricci scalar inside the star means that $\rho+p$ could be negative (remember that $R''$ relates $\rho+p$ at leading order by Eq.~\eqref{JGRG13} or \eqref{FRN12}), which indicates the violation of energy conditions. 
Hence there is no other way than connecting smoothly and monotonically to the outer asymptotic Schwarzschild solution.

Furthermore, in the $F(R)$ gravity with the non-integer power of $R$, the chameleon mechanism~\cite{Khoury:2003rn,Khoury:2003aq,Hu:2007nk,Cognola:2007zu} works well, and the deviation from the Einstein gravity outside the star becomes small. 
To see how the chameleon mechanism work in this model, let us consider the effective mass of the scalar mode.
For general $F(R)$ model, taking the trace of Eq.~\eqref{JGRG13} gives
\begin{align}
\label{TK001}
\Box F_R
= \frac{1}{3} \left[ 2F(R) - RF_R(R) + \kappa^2 T^{\mu}_{\ \mu} \right]\, .
\end{align}
Here, the scalar mode $\Phi$ is defined as $\Phi \equiv F_{R}(R)$, and thus $R = R(\Phi)$.
Reading the second derivative of the effective potential as the square of effective mass (for derivations, see Ref.~\cite{Katsuragawa:2019uto}), one finds
\begin{align}
\label{TK002}
m^{2}_{\Phi} 
& \equiv \frac{1}{3}\frac{F_{R}(R) - RF_{RR}(R)}{F_{RR}(R)}
\end{align}
Applying the mass formula \eqref{TK002} to Eq.~\eqref{Nsurface7}, one finds the effective mass approximately evaluated as follows:
\begin{align}
\label{TK003}
    m^{2}_{\Phi} 
    & \approx \frac{1}{3} \frac{R_{0}}{\beta F_1} \left[ F_0 \left( \frac{R}{R_0} \right)^{1-\beta}+ (1-\beta) F_1 \frac{R}{R_0} \right] \, .
\end{align}
Eq.~\eqref{TK003} shows the effective mass depends on the curvature (or equivalently, the energy density), and the first term is dominant around the surface of the star for small $F_{1}$ because $0<1-\beta<1$.
The scalar mode becomes more massive inside the star, although it is expected to be almost massless outside the star when $\Lambda$ is cast as the dark energy.
Thus, the chameleon mechanism works, and it can screen the scalar mode around the star in the non-integer power model.
At last, we emphasize that in the above argument, we only need the behavior of the surface, and therefore we need not the boundary conditions on the core of the compact star and spatial infinity.

\subsection{Mass-polytrope case}

As another example, we also consider the mass-polytrope EOS:
\begin{align}
\label{MassPolytropicEOS}
\rho = \rho_{m} + n_{m} p \, , \quad p = K_{m} \rho_{m}^{1+\frac{1}{n_{m}}} 
\, ,
\end{align}
where $\rho_{m}$ is the rest-mass energy density with $K_{m}$ and $n_{m}$ being constants.
In a manner same as the analysis for the energy-polytrope, Eq.~\eqref{FRN3} suggests
\begin{align}
\nu (r) = \Bar{\nu}_0 {\textcolor{black}{-}} \ln \left[ 1 - \Tilde{n}_{m} \Tilde{K}^{-1}_{m} p^{-\frac{1}{\Tilde{n}_{m}}}\right] \, .
\end{align}
Here 
\begin{align}
\tilde{K}_{m} \equiv K_{m}^{-\frac{1}{1+\frac{1}{n_{m}}}} \, , \quad
\tilde{n}_{m} \equiv \frac{1}{\frac{1}{1+\frac{1}{n_{m}}} - 1} 
= - 1 - n_{m} \, ,
\end{align}
and $\Bar{\nu}_0$ is a constant.

Assuming the profile of pressure near the surface as in Eq.~\eqref{ProG1B}, the expanded 
\begin{align}
\nu (r) \sim 
\bar{\nu}_0 {\textcolor{black}{-}} \ln \left \{ 1 + (n_{m}+1) \bar{\nu}_1 \left( 1 - \frac{r}{R_s} \right)^{\frac{m}{1+n_{m}}} 
\left[ 1 + p_1 \left( 1 - \frac{r}{R_s} \right) + p_2 \left( 1 - \frac{r}{R_s} \right)^2 + \cdots \right]^{\frac{1}{1+n_{m}}}
\right \} \, ,
\end{align}
where 
\begin{align}
\bar{\nu}_1 \equiv K_{m}^{\frac{1}{1+\frac{1}{n_{m}}}} p_0^{\frac{1}{1+n_{m}}} \, .
\end{align}
The conditions for the continuity of $\nu (r)$ and $\nu'(r)$ lead to $\frac{1}{2} \ln \left( 1 - \frac{2M}{R_s} \right) = \bar{\nu}_0$ and
\begin{align}
\frac{\frac{M}{R_s^2}}{1 - \frac{2M}{R_s}}
=& 
\left\{ \frac{m\bar{\nu}_1}{R_s}\left( 1 - \frac{r}{R_s} \right)^{\frac{m}{1+n_{m}}-1} 
\left[ 1 + p_1 \left( 1 - \frac{r}{R_s} \right) + p_2 \left( 1 - \frac{r}{R_s} \right)^2 + \cdots \right]^{\frac{1}{1+n_{m}}} \right. \nn
& \left. + \frac{\bar{\nu}_1}{R_s} \left( 1 - \frac{r}{R_s} \right)^{\frac{m}{1+n_{m}}} 
\left[ 1 + p_1 \left( 1 - \frac{r}{R_s} \right) + p_2 \left( 1 - \frac{r}{R_s} \right)^2 + \cdots \right]^{\frac{1}{1+n_{m}} - 1}
\left[ p_1 + 2 p_2 \left( 1 - \frac{r}{R_s} \right) + \cdots \right] \right\} \nn
& \left. \times \left\{ 1 + (n_{m}+1) \bar{\nu}_1 \left( 1 - \frac{r}{R_s} \right)^{\frac{m}{1+n_{m}}} 
\left[ 1 + p_1 \left( 1 - \frac{r}{R_s} \right) + p_2 \left( 1 - \frac{r}{R_s} \right)^2 + \cdots \right]^{\frac{1}{1+n_{m}}}\right\}^{-1} \right|_{r=R_s}\, .
\end{align}

The finiteness and non-vanishing condition for $\nu'(r)$ gives
\begin{align}
\label{Finiteseness&NonVanishing_masspoly}
m=n_{m}+1 \, ,
\end{align}
and the argument similar with the derivation of \eqref{ProG7B} leads to
\begin{align}
\label{massMR}
\frac{M}{R_s}=\frac{\left(n_{m}+1 \right) K_{m}^{\frac{1}{1+\frac{1}{n_{m}}}} p_0^{\frac{1}{1+n_{m}}}}
{1{\textcolor{black}{+}}2\left(n_{m}+1 \right) K_{m}^{\frac{1}{1+\frac{1}{n_{m}}}} p_0^{\frac{1}{1+n_{m}}}} 
= \frac{1}{2} \left( 1 - \frac{1}{1{\textcolor{black}{+}}2\left(n_{m}+1 \right) K_{m}^{\frac{1}{1\frac{1}{n_{m}}}} p_0^{\frac{1}{1+n_{m}}}} \right) \, .
\end{align}
\textcolor{black}{
Therefore Eq.~\eqref{massMR} gives the constraint on the compactness of the star written in the mass $M$ and the radius $R_s$, which is identical with Eq.~\eqref{FRsrfc0} in the energy-polytrope case. 
}

With Eq.~\eqref{Finiteseness&NonVanishing_masspoly}, the profile of pressure and its derivatives can be rewritten as below:
\begin{align}
\nu (r) \sim&  
\bar{\nu}_0 {\textcolor{black}{-}} \ln \left \{ 1 + (n_{m}+1) \bar{\nu}_1 \left( 1 - \frac{r}{R_s} \right)
\left[ 1 + p_1 \left( 1 - \frac{r}{R_s} \right) + p_2 \left( 1 - \frac{r}{R_s} \right)^2 + \cdots \right]^{\frac{1}{1+n_{m}}}
\right\} \, . \\
\nu' (r) \sim 
&\left\{ \frac{(n_{m}+1)\bar{\nu}_1}{R_s}
\left[ 1 + p_1 \left( 1 - \frac{r}{R_s} \right) + p_2 \left( 1 - \frac{r}{R_s} \right)^2 + \cdots \right]^{\frac{1}{1+n_{m}}} \right. \nn
& \left. + \frac{\bar{\nu}_1}{R_s} \left( 1 - \frac{r}{R_s} \right)
\left[1 + p_1 \left( 1 - \frac{r}{R_s} \right) + p_2 \left( 1 - \frac{r}{R_s} \right)^2 + \cdots \right]^{\frac{1}{1+n_{m}} - 1}
\left[ p_1 + 2 p_2 \left( 1 - \frac{r}{R_s} \right) + \cdots \right] \right\} \nn
& \times \left\{ 1 + (n_{m}+1) \bar{\nu}_1 \left( 1 - \frac{r}{R_s} \right)
\left[ 1 + p_1 \left( 1 - \frac{r}{R_s} \right) + p_2 \left( 1 - \frac{r}{R_s} \right)^2 + \cdots \right]^{\frac{1}{1+n_{m}}}\right\}^{-1} 
\, , \\
\nu'' (r) \sim 
& {\textcolor{black}{-}}\left\{ \frac{2\bar{\nu}_1}{R_s^2} 
\left[ 1 + p_1 \left( 1 - \frac{r}{R_s} \right) + p_2 \left( 1 - \frac{r}{R_s} \right)^2 + \cdots \right]^{\frac{1}{1+n_{m}} - 1}
\left[ p_1 + 2 p_2 \left( 1 - \frac{r}{R_s} \right) + \cdots \right] \right. \nn
& - \frac{n_{m} \bar{\nu}_1}{\left( 1 + n_{m} \right) R_s^2} \left( 1 - \frac{r}{R_s} \right) 
\left[1 + p_1 \left( 1 - \frac{r}{R_s} \right) + p_2 \left( 1 - \frac{r}{R_s} \right)^2 + \cdots \right]^{\frac{1}{1+n_{m}} - 2}
\left[ p_1 + 2 p_2 \left( 1 - \frac{r}{R_s} \right) + \cdots \right]^2 \nn 
& \left. + \frac{2\bar{\nu}_1 p_2}{R_s^2} \left( 1 - \frac{r}{R_s} \right) 
\left[ 1 + p_1 \left( 1 - \frac{r}{R_s} \right) + p_2 \left( 1 - \frac{r}{R_s} \right)^2 + \cdots \right]^{\frac{1}{1+n_{m}} - 1}
+ \cdots \right\} \nn
& \times \left\{ 1 + (n_{m}+1) \bar{\nu}_1 \left( 1 - \frac{r}{R_s} \right)
\left[ 1 + p_1 \left( 1 - \frac{r}{R_s} \right) + p_2 \left( 1 - \frac{r}{R_s} \right)^2 + \cdots \right]^{\frac{1}{1+n_{m}}}\right\}^{-1} \nn
& {\textcolor{black}{+}} \frac{1}{1 + n_{m}} \left\{ \frac{\left( 1 + n_{m} \right) \bar{\nu}_1}{R_s} \left[ 1 + p_1 \left( 1 - \frac{r}{R_s} \right) 
+ p_2 \left( 1 - \frac{r}{R_s} \right)^2 + \cdots \right]^{\frac{1}{1+n_{m}}} \right. \nn
& \left. + \frac{\bar{\nu}_1}{R_s} \left( 1 - \frac{r}{R_s} \right) 
\left[ 1 + p_1 \left( 1 - \frac{r}{R_s} \right) + p_2 \left( 1 - \frac{r}{R_s} \right)^2 + \cdots \right]^{\frac{1}{1+n_{m}} - 1}
\left[ p_1 + 2 p_2 \left( 1 - \frac{r}{R_s} \right) + \cdots \right] \right\}^2 \nn
& \times \left\{ 1 + (n_{m}+1) \bar{\nu}_1 \left( 1 - \frac{r}{R_s} \right)
\left[ 1 + p_1 \left( 1 - \frac{r}{R_s} \right) + p_2 \left( 1 - \frac{r}{R_s} \right)^2 + \cdots \right]^{\frac{1}{1+n_{m}}}\right\}^{-2} \, . 
\label{NuDerivsOnSurface}
\end{align}
$\nu''(r)$ should be continuous on the surface due to the non-singularity of the equation of motion. 
From Eqs.~\eqref{Sch1} and \eqref{NuDerivsOnSurface},
we obtain 
\begin{align}
 - \frac{\frac{2M}{R_s^3}}{1 - \frac{2M}{R_s}} 
 - \frac{\frac{2M^2}{R_s^4}}{\left(1 - \frac{2M}{R_s}\right)^2}
={\textcolor{black}{-}}\frac{2\bar{\nu}_1 p_1}{R_s^2} {\textcolor{black}{+}} \frac{\left( 1 + n_{m} \right) \bar{\nu}_1^2}{R_s^2} \, ,
\end{align}
which determines $p_1$
\begin{align}
p_1 = \frac{\left( 1 + n_{m} \right) \bar{\nu}_1}{2}
 {\textcolor{black}{+}} \frac{1}{\bar{\nu}_1} \left[ 
 \frac{\frac{M}{R_s}}{1 - \frac{2M}{R_s}} +\frac{\frac{M^2}{R_s^2}}{\left(1 - \frac{2M}{R_s}\right)^2} \right]
\, .
\end{align}
The continuity for $\nu'''(r)$ is also demanded, and $p_2$ in \eqref{ProG1B} can be calculated in the same way. 

The arguments on the continuity conditions for the scalar curvature and the asymptotic behavior of $F(R)$ in Eqs.~\eqref{sc1}-\eqref{FRp3} do not depend on the EOS, and thus they also hold in the mass-polytrope case. 
The EOS Eqs.~\eqref{MassPolytropicEOS} and pressure profile Eq.~\eqref{ProG1B} read 
\begin{align}
p \sim p_0 \left( 1 - \frac{r}{R_s} \right)^{n_{m}+1}\, , \quad 
\rho \sim \Tilde{K}_m p_0^\frac{n_{m}}{n_{m}+1} \left( 1 - \frac{r}{R_s} \right)^{n_{m}}
+ n_{m} p_0 \left( 1 - \frac{r}{R_s} \right)^{n_{m}+1}
\sim \left( 1 - \frac{r}{R_s} \right)^{n_{m}}\, .
\end{align}
As Eq.~\eqref{prho} relations above suggest
\begin{align}
    \rho + p \sim \left( 1 - \frac{r}{R_s} \right)^{n_{m}}\, ,
\end{align}
and Eq.~\eqref{FRp3} gives
\begin{align}
\beta = \frac{n_{m}+2}{\alpha} \, .
\end{align}
Therefore, the same conclusion as in the energy-polytrope case Eq.~\eqref{Nsurface7} follows in the mass-polytrope case.

\section{Summary and Discussions \label{SecV}}

Investigating the spherically symmetric and static space-time in the $F(R)$ gravity, 
we have found the differential equations for $F_R(r)= \left. \frac{dF(R)}{dR}\right|_{R=R(r)}$
based on a new formalization of the TOV equations.
The relation between the mass $M$ and the radius $R_s$ of the compact star can be determined if the energy (mass) density profile $\rho=\rho(r)$ is given. 
When we assume a certain EOS, we can obtain the pressure $p$ as a function of $r$, $p=p\left(\rho\left(r\right)\right)$. 
The profile $\rho=\rho(r)$ also determines the $r$ dependence of $\nu=\nu(r)$ and  the differential equation for $F_R(r)$ is given in a closed form. 
By solving the differential equation, 
we find the $r$ dependence of  $F_R(r)$, which gives $\lambda(r)$, and also $R(r)$, as functions of $r$. 
Then by combining the $r$ dependence of $F_R(r)$ and $R(r)$, 
the functional form of $F_R \left(R\right)$ and therefore $F \left(R\right)$ can be determined.

Furthermore, we have solved the equations perturbatively for the energy polytrope and mass polytrope. 
This paper claims that the correction to the Einstein gravity can take the form of the non-integer power of $R$,
which can be found by solving the above differential equations near and inside the surface of the compact star. 
The non-integer power shows up because we require that the first derivative $R'$ of $R$ be continuous 
with respect to $r$, which follows that in the case of the $F(R)$ gravity, the equations include the second-order derivatives of the curvatures as given in Eqs.~\eqref{FRN4} and \eqref{FRN6}. 
If there is a discontinuity in $R'$, 
$R''$ includes the singularity which behaves as the delta function and 
Eqs.~\eqref{FRN4} and \eqref{FRN6} tell that the energy density and the pressure 
also must have the delta function singularity corresponding to the shell. 

Although our results follow from the analytic EOS, it provides us with significant guidelines for model-building in $F(R)$ gravity theory.
From the viewpoint of cosmology,
phenomenological models of $F(R)$ gravity has been proposed to explain the accelerated expansion of the present Universe,
which includes the low-energy correction to the Einstein gravity.
However, those models suffer from the curvature singularities~\cite{Frolov:2008uf, Bamba:2011sm} at a finite field value,
which requires the higher-curvature corrections to push the singularity away to the infinite value of the field.
In a way similar to the above,
one can see that our results constrain the model of $F(R)$ gravity from the viewpoint of astrophysics,
which suggests that $F(R)$ gravity can include the non-integer power of the curvature in its action.
It is remarkable that the correction term to the Einstein gravity is not necessarily the integer-power form, such as $R^2$, but can be more general.
Our analysis shows that consistency with the exterior solution of the compact star can give us constraints on the $F(R)$ function.
It is non-trivial if the Schwarzschild space-time is a proper boundary condition for the compact star in arbitrary $F(R)$ gravity, which may indicate the implicit problems in the existing research.
We will revisit the existing studies and address the boundary condition problem in future works.

We make one more comment on our consequence from the viewpoint of field theory.
It might look strange that the non-integer power of the scalar curvature appears in the action, but it is not so when we regard the $F(R)$ gravity as a low-energy effective theory of gravitation. 
Just for example, as a model in the high-energy region, we may consider the Brans-Dicke or dilaton type action with a scalar field $\phi$ included: 
\begin{align}
\label{Sphi1}
S_\phi = \int d^4 x \sqrt{-g} \left( \phi R - \frac{\omega(\phi)}{2} \partial_\mu \phi \partial^\mu \phi - V(\phi) \right) \, .
\end{align}
When we consider the low-energy effective action, we may ignore the kinetic term $- \frac{\omega(\phi0)}{2} \partial_\mu \phi \partial^\mu \phi$. 
Just for simplicity, we assume the potential $V(\phi)$ is given by a power of $\phi$: $V(\phi) = V_0 \phi^n$ with constants $V_0$ and $n$. 
By ignoring the kinetic term, the equation given by the variation with respect to $\phi$ has the following form, 
\begin{align}
\label{Sphi2}
R=n V_0 \phi^{n-1} \, .
\end{align}
Rewriting $\phi$ in terms of $R$ in Eq.~\eqref{Sphi1}, we obtain the following effective action,
\begin{align}
\label{Sphi3}
S_{\phi\, \mathrm{eff}}= \int d^4 x \sqrt{-g} \left( 1 - \frac{1}{n} \right) \frac{R^{\frac{1}{n-1} +1}}{\left( nV_0\right)^{\frac{1}{n-1}}} \, ,
\end{align}
which shows the fractional power or non-integer of the scalar curvature $R$.

\section*{Acknowledgments} 

This work was supported by JSPS Grant-in-Aid for Scientific 
Research (C) No. 18K03615 (S.~N.). 
T.K is supported by the start-up grant by Central China Normal University.

\appendix

\section{Geometrical Quantities}

In this work, the conventions and definitions of the connections and the curvatures are following:
\begin{align}
&
\Gamma^\eta_{\mu\lambda} =
\frac{1}{2}g^{\eta\nu}\left(
g_{\mu\nu,\lambda} + g_{\lambda\nu,\mu} - g_{\mu\lambda,\nu} 
\right)
\, , 
\quad 
R^\lambda_{\ \mu\rho\nu} =
 -\Gamma^\lambda_{\mu\rho,\nu}
+ \Gamma^\lambda_{\mu\nu,\rho}
 - \Gamma^\eta_{\mu\rho}\Gamma^\lambda_{\nu\eta}
+ \Gamma^\eta_{\mu\nu}\Gamma^\lambda_{\rho\eta} 
\, , \nn 
&
R_{\mu\nu} =
R^\lambda_{\ \mu\lambda\nu} 
\, , 
\quad 
R =
g^{\mu\nu}R_{\mu\nu} 
\, .
\end{align}
For the metric~\eqref{GBiv}, 
we define the metric $\bar{g}_{ij}$ of the unit 2-sphere by 
$\sum_{i,j=1,2} \bar{g}_{ij} dx^i dx^j = d\theta^2 + \sin^2\theta 
\, d\phi^2$. 
The non-vanishing components of the connection coefficients are, 
\begin{align}
\label{GBv0S}
\begin{split}
&\Gamma^r_{tt} = \e^{-2(\lambda - \nu)}\nu' \, ,
\quad \Gamma^t_{tr}=\Gamma^t_{rt}=\nu'\, , \quad 
\Gamma^r_{rr}=\lambda'
\, , \\
&\Gamma^i_{jk} = \bar{\Gamma} ^i_{jk}\, ,\quad 
\Gamma^r_{ij}=-\e^{-2\lambda}r \bar{g}_{ij} \, ,
\quad \Gamma^i_{rj}=\Gamma^i_{jr}=\frac{1}{r} \, \delta^i_{\ j}
\, , \quad     
\end{split}
\end{align}
And, the non-vanishing components of the curvature tensors are, 
\begin{align}
\label{curvaturesS}
R_{rtrt} =&\, \e^{2\nu}\left[ \nu'' + \left(\nu' - \lambda'\right)\nu' \right] 
\, ,\quad 
R_{titj} = r\nu'\e^{2(\nu - \lambda)} \bar{g}_{ij} 
\, , 
\nn
R_{rirj} =& \, \lambda' r \bar{ g}_{ij} \, , \quad 
R_{ijkl} = \left( 1 - \e^{-2\lambda}\right) r^2
\left(\bar{g}_{ik} \bar{g}_{jl} - \bar{g}_{il} \bar{g}_{jk} \right)
\, , 
\nn
R_{tt} =&\, \e^{2\left(\nu - \lambda\right)} \left[
\nu'' + \left(\nu' - \lambda'\right)\nu' + \frac{2\nu'}{r}\right] 
\, , 
\quad  
R_{rr} =
 - \left[ \nu'' + \left(\nu' - \lambda'\right)\nu' \right] + \frac{2 \lambda'}{r} 
\, , 
\nn
R_{ij} =& \left[ 1 + \left\{ - 1 - r \left(\nu' - \lambda' \right)\right\}
\e^{-2\lambda}\right] \bar{g}_{ij}
\, , 
\quad 
R=
\e^{-2\lambda}\left[ - 2\nu'' - 2\left(\nu'
 - \lambda'\right)\nu' - \frac{4\left(\nu'
 - \lambda'\right)}{r} + \frac{2\e^{2\lambda} - 2}{r^2} \right] 
 \, .
\end{align}


\end{document}